\documentstyle[prd,preprint,tighten,aps,amssymb,newlfont,epsfig]{revtex}
\begin{document}
\draft
\preprint{
\begin{tabular}{r}
   UWThPh-1999-54
\\ PRL-TH-1999
\end{tabular}
}

\title{FIELD-THEORETICAL TREATMENT\\ OF NEUTRINO OSCILLATIONS:\\
THE STRENGTH OF THE CANONICAL OSCILLATION FORMULA\thanks{Talk given by
W. Grimus at \textit{Neutrino Mixing}, Meeting in Honour of Samoil
Bilenky's 70th Birthday, Torino, March 25--27, 1999.}}

\author{W. GRIMUS}

\address{Institute for Theoretical Physics, University of Vienna,
Boltzmanngasse 5,\\ A-1090 Vienna, Austria\\
E-mail: grimus@doppler.thp.univie.ac.at}

\author{S. MOHANTY AND P. STOCKINGER}

\address{Theory Group, Physical Research Laboratory,
Ahmedabad - 380 009, India\\
E-mail: mohanty@prl.ernet.in and stocki@prl.ernet.in}  

\maketitle

\begin{abstract}
We discuss conceptual aspects of neutrino oscillations with the main 
emphasis on the field-theoretical approach. 
This approach includes the neutrino source and
detector processes and allows to obtain the neutrino transition or survival
probabilities as cross sections derived from the Feynman diagram
of the combined source -- detection process. In this context, 
the neutrinos which are supposed to oscillate appear as
propagators of the neutrino mass eigenfields, connecting the
source and detection processes.
We consider also the question why the canonical neutrino oscillation
formula is so robust against corrections and discuss the nature of the
oscillating neutrino state emerging in the field-theoretical approach.
\end{abstract}

\newpage

\section{Introduction}

\subsection{History}

Neutrino oscillations play a central role in neutrino
physics. The idea of neutrino oscillations in analogy with 
$K^0 \bar K^0$ oscillations was first discussed by B. Pontecorvo in
the late fifties \cite{pon1}. The concept of 2-neutrino flavour
mixing~\cite{maki} (1962)
and the confirmation of the existence of 
different neutrino flavours~\cite{brookhaven} $\nu_e$ and $\nu_\mu$ 
in the same year opened the way to neutrino flavour oscillations \cite{pon2}.
For a general number of neutrinos, neutrino mixing is denoted as
\begin{equation}\label{mixing}
\nu_{L\alpha} = \sum_j U_{\alpha j} \nu_{Lj}
\quad \mbox{with} \quad
\alpha = e, \mu, \tau, \ldots,
\quad
j = 1,2,3, \ldots
\end{equation}
labelling neutrinos flavours (types) and mass eigenfields,
respectively. 

The first theory of 2-neutrino oscillations was developed by Gribov
and Pontecorvo \cite{gribov}. The 2-neutrino oscillation
probabilities in the form as they are used nowadays were formulated by
Bilenky and Pontecorvo in Ref.~\cite{bilenky} and the extension to an
arbitrary number of neutrinos (with $U$ real)
can be found in Refs.~\cite{generalP,pon}.\footnote{For a detailed
account of the history of neutrino oscillations see the contribution
of S.M. Bilenky to this meeting (hep-ph/9908335).} 
The most general form of
transition ($\alpha \neq \beta$) and survival ($\alpha = \beta$) 
probabilities (see, e.g., Refs.~\cite{hosek,BP87}) reads
\begin{equation}
P_{\nu_\alpha\to\nu_\beta} (L/E_\nu) =
\left| \sum_j U_{\beta j} U^*_{\alpha j} 
\exp \left( -i \frac{m^2_j L}{2E_\nu} \right) \right|^2 \,,
\label{P}
\end{equation}
where $U$ denotes the unitary mixing matrix, $L$ the distance between
source and detector and $E_\nu$ the neutrino energy. The neutrino masses
$m_j$ are associated with the mass eigenfields $\nu_j$.
Note that the above formula assumes ultra-relativistic neutrinos.
With the ordering $m_1 \leq m_2 \leq \ldots$, Eq.~(\ref{P}) is rewritten as
\begin{eqnarray}
\lefteqn{P_{\nu_\alpha\to\nu_\beta} (L/E_\nu) =} \nonumber \\
& & \sum_j |U_{\beta j}|^2 |U_{\alpha j}|^2 + 
2 \sum_{k>j} \, \mbox{Re} \, 
\left[U_{\beta j}^* U_{\alpha j} U_{\beta k} U_{\alpha k}^* 
\exp \left( -i 2\pi L/L^\mathrm{osc}_{kj} \right) \right] \,,
\label{PP}
\end{eqnarray}
where the oscillation lengths are defined by~\cite{pon}
\begin{equation}\label{Losc}
L^\mathrm{osc}_{kj} = 4\pi \frac{E_\nu}{\Delta m^2_{kj}} \,.
\end{equation}

It was a very important discovery that the oscillation pattern as described by
Eqs.~(\ref{P}) and (\ref{PP}) can get significantly modified
by the effect of coherent forward scattering in 
background matter \cite{MSW}. Otherwise,
the canonical oscillation formula (\ref{P}) or its equivalent form
(\ref{PP}) has resisted -- at least in terrestrial experiments -- 
all attempts to find non-negligible corrections.
The strong evidence for neutrino oscillations found in atmospheric neutrino
experiments~\cite{SK-evidence} (for reviews on neutrino oscillations
see, e.g., Ref.~\cite{reviews}) constitutes a major achievement and
progress in neutrino physics. If in addition to the solar neutrino
evidence for neutrino oscillations also the LSND evidence will be
confirmed by future experiments, then neutrino physics is enriched by
a sterile neutrino (or sterile neutrinos), a notion coined by 
Pontecorvo in Ref.~\cite{pon2}. In this case, in the summation in
Eq.~(\ref{mixing}) there will be sterile neutrino types in addition to
the flavours.
For a review of the physics of sterile
neutrinos see Ref.~\cite{mohapatra}.

\subsection{Conceptual questions}
\label{concept}

It has been indicated in several publications that the 
standard derivation of Eq.~(\ref{P}) as found, e.g., in
Ref.~\cite{pon} raises a number of conceptual questions
(see, e.g., Ref.~\cite{rich} for a clear exposition). Let us consider
some of these questions for a moment:
\begin{enumerate}
\item
Under which conditions is Eq.~(\ref{P}) valid?
\item
Do the neutrino mass eigenstates $|\nu_j \rangle$ have definite
momenta or definite energies? Or are both smeared out? Note that in
the usual derivation it is assumed that all $|\nu_j \rangle$ have the
same momenta $p$ and thus energies $E_j = \sqrt{m_j^2 + p^2}$. One
considers their time evolution given by $\exp (-iE_jt)$ and finally
makes the replacement $t \to L/c$ ($c$ is the velocity of light). The
last step is necessary to make contact with realistic experiments where
the distance $L$ between source
and detector is known, but no measurement of the time $t$ can be
made \cite{lipkin}.
\item
Why is it allowed to make the replacement $t \to L/c$ when
wave functions with definite momentum have infinite extent in space?
\item
Why not making the replacement $t \to cL/v_j$ where $v_j = p/E_j$ is
the velocity of the neutrino with mass $m_j$?
\item
Evidently, the Fock space of neutrinos with definite mass exists. Is
this also the case for flavour neutrino states as sometimes assumed in
the literature?
\end{enumerate}

One aspect of the limitation of Eq.~(\ref{P}) is treated in terms
of a coherence length \cite{nussinov,kim}.
Some of the above questions are answered in the framework of 
the wave packet approach~\cite{kayser,giunti2}
(see also the review~\cite{zralek} where a list of references can be
found), where the neutrino momentum is smeared out,
however, the size and form of wave packet is not determined in
this approach and remains a subject to reasonable estimates.
Let us shortly mention the answers to the above questions offered by
the wave packet approach:
\newcounter{ad}
\begin{list}{Ad \arabic{ad}:}{}
\usecounter{ad}
\item
One necessary condition for validity of the canonical formula
(\ref{P}) is formulated using the notion of a coherence length.
Any coherence length in neutrino oscillations~\cite{nussinov} 
is related to the oscillation length (\ref{Losc}) by~\cite{KNW96}
\begin{equation}\label{Lcoh}
L^\mathrm{coh}_{kj} = 
\frac{\bar E_\nu}{\Delta E_\nu} \, L^\mathrm{osc}_{kj}\,,
\end{equation}
where $\bar E_\nu$ is the mean neutrino energy and $\Delta E_\nu$ is the
energy spread of the neutrino beam. If $L \gg L^\mathrm{coh}_{kj}$
holds, the term with $\exp (-i2\pi L/L^\mathrm{coh}_{kj})$ in
Eq.~(\ref{PP}) is suppressed.
\item
Obviously, smearing out the neutrino momentum also smears out the
neutrino energies.
\item
The wave packet approach deals with the time variable either by
putting~\cite{kayser} $t = L/c$ or by averaging~\cite{giunti2} over $t$.
\item
Such a replacement in the standard derivation of
Eq.~(\ref{P}) is wrong because it leads to 
$E_j t_j = E_j^2L/p = pL + m_j^2 L/p$ and consequently misses the
factor 1/2 in the exponents \cite{lipkin,grossman}.
\item
In Ref.~\cite{giunti1} arguments are given that a Fock space of flavour
neutrinos does not exist.
\end{list}

\subsection{The field-theoretical approach}

The idea has been put forward to include the neutrino production and
detection processes into the consideration of neutrino
oscillations \cite{rich}. Such an approach can be realized with 
\emph{quantum mechanics} --
in which case the neutrinos with definite mass are
unobserved intermediate states between the source and detection
processes~\cite{rich} -- or with \emph{quantum field theory} where the
massive neutrinos are represented by inner lines in a big Feynman
diagram depicting the combined source -- detection
process \cite{GS96,GSM99}. In the following we will discuss the
field-theoretical treatment. The aims and hopes of such an approach
are the following: 
\begin{enumerate}
\renewcommand{\labelenumi}{\Alph{enumi}.}
\item
The elimination of the arbitrariness associated
with the wave packet approach,
\item
the description of neutrino
oscillations by means of the particles in neutrino production and detection 
which are really manipulated in an experiment, 
\item
a more complete and
realistic description in order to find possible limitation of formula
(\ref{P}) in specific experimental situations. 
\end{enumerate}
Note that in this approach the question of a Fock space for flavour
states has no relevance.

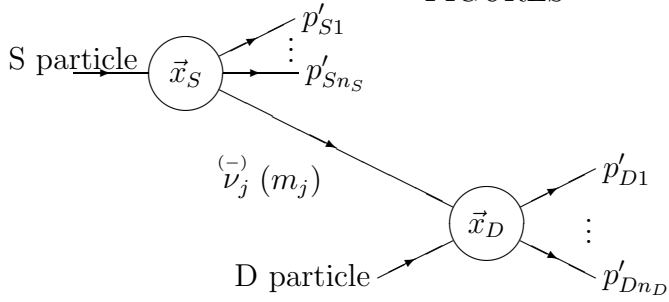
\begin{figure}[t]
\setlength{\unitlength}{5mm}
\begin{picture}(12,8)(-4,1)
\put(0,8){\vector(1,0){1}} \put(1,8){\line(1,0){1}}
\put(3,8){\circle{2}} \put(3,8){\makebox(0,0){$\vec{x}_S$}}
\put(3.8944,8.4472){\vector(2,1){1}}
\put(4.8944,8.9472){\line(2,1){1}}
\put(4,8){\vector(1,0){1}} \put(5,8){\line(1,0){1}}
\put(3.8944,7.5528){\vector(2,-1){3.1056}}
\put(7,6){\line(2,-1){3.1056}}
\put(11,4){\circle{2}} \put(11,4){\makebox(0,0){$\vec{x}_D$}}
\put(11.8944,4.4472){\vector(2,1){1}}
\put(12.8944,4.9472){\line(2,1){1}}
\put(11.8944,3.5528){\vector(2,-1){1}}
\put(12.8944,3.0528){\line(2,-1){1}}
\put(8.1056,2.5528){\vector(2,1){1}}
\put(9.1056,3.0528){\line(2,1){1}}
%
\put(0,8.1){\makebox(0,0)[b]{S particle}}
\put(6.0944,9.4472){\makebox(0,0)[l]{$p'_{S1}$}}
\put(6.2,8){\makebox(0,0)[l]{$p'_{S{n_S}}$}}
\put(5.8,8.3){\makebox(0,0)[b]{$\vdots$}}
\put(6.9,5.9){\makebox(0,0)[tr]{
$\stackrel{\scriptscriptstyle (-)}{\nu}_{\hskip-3pt j}(m_j)$
}}
\put(7.9056,2.5528){\makebox(0,0)[r]{D particle}}
\put(14.0944,5.4472){\makebox(0,0)[l]{$p'_{D1}$}}
\put(14.0944,2.5528){\makebox(0,0)[l]{$p'_{D{n_D}}$}}
\put(13.6,4){\makebox(0,0)[l]{$\vdots$}}
\end{picture}
\caption{\label{DAR} Feynman diagram for decay at rest (DAR) of
the neutrino source particle. The source (S) and detector (D) processes
are symbolized by the circles. The labels $\vec{x}_S$ and
$\vec{x}_D$ represent the coordinates where the wave functions
of the source and detector particles are peaked, respectively.
We have also indicated the $n_S$ ($n_D$) momenta of the final particles
originating from the source (detector) process and the neutrino
propagator of the neutrino field with mass $m_j$.}
\end{figure}

\begin{figure}[t]
\setlength{\unitlength}{5mm}
\begin{picture}(12,10)(-4,-2)
\put(0,8){\vector(1,0){1}} \put(1,8){\line(1,0){1}}
\put(3,8){\circle{2}} \put(3,8){\makebox(0,0){$\vec{x}_T$}}
\put(3.8944,8.4472){\vector(2,1){1}}
\put(4.8944,8.9472){\line(2,1){1}}
\put(4,8){\vector(1,0){1}} \put(5,8){\line(1,0){1}}
\put(0.2929,5.2929){\vector(1,1){1}}
\put(1.2929,6.2929){\line(1,1){1}}
\put(3.8944,7.5528){\vector(2,-1){2.1056}}
\put(6,6.5){\line(2,-1){2}}
\put(8,5.5){\vector(1,-2){1}}
\put(9,3.5){\line(1,-2){0.5528}}
\put(8,5.5){\vector(2,1){1}} \put(9,6){\line(2,1){1}}
\put(8,5.5){\vector(1,0){1}} \put(9,5.5){\line(1,0){1}}
\put(10,1.5){\circle{2}} \put(10,1.5){\makebox(0,0){$\vec{x}_D$}}
\put(10.8944,1.9472){\vector(2,1){1}}
\put(11.8944,2.4472){\line(2,1){1}}
\put(10.8944,1.0528){\vector(2,-1){1}}
\put(11.8944,0.5528){\line(2,-1){1}}
\put(7.1056,0.0528){\vector(2,1){1}}
\put(8.1056,0.5528){\line(2,1){1}}
%
\put(0.8,8.1){\makebox(0,0)[b]{proton}}
\put(0.8,7.8){\makebox(0,0)[t]{$\vec{P}$}}
\put(6.0944,9.4472){\makebox(0,0)[l]{$p'_{T1}$}}
\put(6.2,8){\makebox(0,0)[l]{$p'_{T{n_T}}$}}
\put(5.8,8.3){\makebox(0,0)[b]{$\vdots$}}
\put(0.2929,5.1929){\makebox(0,0)[t]{T particle}}
\put(6.2,6.3){\makebox(0,0)[tr]{S particle}}
\put(7.9,5.4){\makebox(0,0)[tr]{$\vec{x}_S$}}
\put(10.2,6.5){\makebox(0,0)[l]{$p'_{S1}$}}
\put(10.2,5.5){\makebox(0,0)[l]{$p'_{S{n_S}}$}}
\put(9.9,5.6){\makebox(0,0)[b]{$\vdots$}}
\put(8.9,3.6){\makebox(0,0)[tr]{
$\stackrel{\scriptscriptstyle (-)}{\nu}_{\hskip-3pt j}(m_j)$
}}
\put(6.9056,0.0528){\makebox(0,0)[r]{D particle}}
\put(13.0944,2.9472){\makebox(0,0)[l]{$p'_{D1}$}}
\put(13.0944,0.0528){\makebox(0,0)[l]{$p'_{D{n_D}}$}}
\put(12.6,1.5){\makebox(0,0)[l]{$\vdots$}}
\end{picture}
\caption{\label{DIF} Feynman diagram for decay in flight (DIF) of
the neutrino source particle which is produced by a proton with
momentum $\vec{P}$ hitting a target (T) particle localized at
$\vec{x}_T$. In addition to the final momenta of DAR there are $n_T$
final momenta originating from the target process.}
\end{figure}
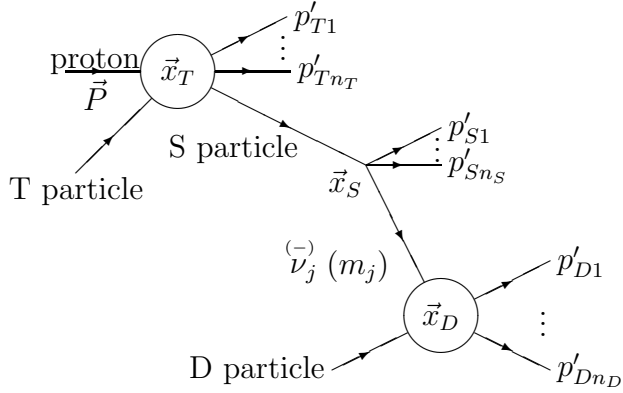

Considering laboratory experiments, there are two typical situations
for neutrino oscillation experiments. The first one is \emph{decay at
rest} (DAR) of the neutrino source. Its corresponding Feynman diagram
is depicted in Fig.~\ref{DAR}. The wave functions of the source and
detector particles are localized (peaked) at $\vec{x}_S$ and
$\vec{x}_D$, respectively. The other situation is \emph{decay in
flight} (DIF) of the neutrino source as represented in
Fig.~\ref{DIF} where it is assumed that a proton hits a target
localized at $\vec{x}_T$. The detector particle sits again at
$\vec{x}_D$ but the source is not localized. In both situations the
distance between source and detection is given by 
$L = |\vec{x}_D-\vec{x}_S|$. Note that in the Feynman diagrams of
Figs.~\ref{DAR} and \ref{DIF} the neutrinos with definite mass occur
as inner lines. In the spirit of our approach, neutrino oscillation
probabilities are proportional to the cross sections derived from the
amplitudes represented by these diagrams.

In the following we will work out the amplitude for neutrino
oscillations as described by Figs.~\ref{DAR} and \ref{DIF} in the
limit of macroscopic $L$, discuss the consequences and compare this
approach with the wave packet approach. Note that
within the field-theoretical approach the notion \emph{coherent}
refers to a summation in the amplitude, whereas \emph{incoherent}
means a summation over squares of amplitudes, i.e., a summation in the
cross section \cite{GSM99}.

\section{Assumptions and the resulting amplitude}
\subsection{The basic assumptions}
\label{assumptions}

The further discussion is based on the following assumptions:
  \begin{enumerate}
  \renewcommand{\labelenumi}{\Roman{enumi}.}
  \item The wave function $\phi_D$ of the detector particle does not spread
    with time, which amounts to
    \begin{equation}\label{psiD}
    \phi(\vec{x},t) = \psi_D(\vec{x}-\vec{x}_D)\, e^{-i E_{DP}t} \,,
    \end{equation}
    where $E_{DP}$ is the sharp energy of the detector particle and
    $\psi_D(\vec{y})$ is peaked at $\vec{y}=\vec{0}$.
  \item The detector is sensitive to momenta (energies) and possibly to
    observables commuting with momenta (charges, spin).
  \item The usual prescription for the calculation of the cross section is
    valid. 
  \end{enumerate}
With the amplitude symbolized by Fig.~\ref{DAR} the
oscillation probability is obtained by
\begin{equation}\label{Pav}
\left\langle 
P_{\stackrel{\scriptscriptstyle (-)}{\nu}_{\hskip-3pt \alpha} 
\to \stackrel{\scriptscriptstyle (-)}{\nu}_{\hskip-3pt \beta}}
\right\rangle_\mathcal{P} \propto
\int dP_S \int_\mathcal{P} \frac{d^3 p_{D1}}{2E_{D1}} \cdots 
\frac{d^3 p_{D{n_D}}}{2E_{D{n_D}}} \: \left| 
\mathcal{A}_{\stackrel{\scriptscriptstyle (-)}{\nu}_{\hskip-3pt \alpha} 
\to \stackrel{\scriptscriptstyle (-)}{\nu}_{\hskip-3pt \beta}}
\right|^ 2 \,.
\end{equation}
In this equation we have indicated the average over some region $\mathcal{P}$
in the phase space of the final particle of the detection process. If no final
particle of the neutrino production process is measured then one has to
integrate over the total phase space of these final states. For DIF
symbolized by Fig.~\ref{DIF} there is an additional integration 
$\int dP_T$ over the final particles in the target process. By definition, at
the source (detector) a neutrino 
$\stackrel{\scriptscriptstyle (-)}{\nu}_{\hskip-3pt \alpha}$ 
($\stackrel{\scriptscriptstyle (-)}{\nu}_{\hskip-3pt \beta}$)
is produced (detected) if there is a charged lepton $\alpha^{\pm}$ 
($\beta^{\pm}$) among the final states.

In perturbation theory with respect to weak interactions, according to the
Feynman diagrams Figs.~\ref{DAR} and \ref{DIF} one has to perform integrations
$\int d^4 x_1$, $\int d^4 x_2$ and $\int d^4 q$ corresponding to the
Hamiltonian densities for neutrino production, detection and the propagators
of the mass eigenfields. These integrations are non-trivial because $\psi_D$
and the source (target) wave functions are not plane waves, but are localized
at $\vec{x}_D$ and $\vec{x}_S$ ($\vec{x}_T$), respectively. 

\subsection{Integration over $d^4 x_2$}

Let us discuss now in detail the integration $\int d^4 x_2$ at
the interaction vertex of the detector process. For definiteness we consider
$\bar{\nu}_\alpha \to \bar{\nu}_\beta$ transitions, which is the typical
situation in experiments with DAR of the neutrino source (see LSND, KARMEN and
reactor experiments). A remark is at order concerning the integration over
$d x^0_2$. We will perform this integration from $-\infty$ to $+\infty$, which
corresponds to initial time $t_i = -\infty$ and final time $t_f = +\infty$ of
the combined source -- detection process. This is certainly only approximately
correct for DAR (see Fig.~\ref{DAR}), because in this case the time evolution
of the source particle should rather be considered starting at a finite time,
after the preparation of the state of the source particle. We will later
comment on this problem in detail. Looking at the neutrino line in
Figs.~\ref{DAR} and \ref{DIF} which connects the neutrino source with the
detector particle, with Assumption II saying that final states are
described by plane waves, we read off the following integral:
\begin{eqnarray}
&& \int d^4 x_2 \exp{\left( i\sum_b p'_{Db} \cdot x_2 \right)}
\int \frac{d^4 q}{(2\pi)^4}\, e^{-iq \cdot (x_1 - x_2)}
\nonumber \\
&& \times 
\sum_j U_{\alpha j}\, \frac{\not\! q + m_j}{q^2 - m_j^2 + i\epsilon} \,
U^*_{\beta j} \, \psi_D(\vec{x}_2 - \vec{x}_D) \, e^{-iE_{DP} t_2}
\nonumber \\
&& = \int d q^0 e^ {-iq^0t} \delta (q^0+E_D) 
\int \frac{d^3 q}{(2\pi)^{3/2}}\, e^{-i\vec{q} \cdot (\vec{x}_D - x_1)}
\nonumber \\
&& \times \sum_j U_{\alpha j} \,
\frac{\not\! q + m_j}{q^2 - m_j^2 + i\epsilon}\, U^*_{\beta j} \,
\widetilde{\psi}_D (\vec{q} + \vec{p}\,'_D) 
\label{x2int}
\end{eqnarray}
with the definitions
\begin{equation}\label{defs}
\vec{p}\,'_D = \sum_{b=1}^{n_D} \vec{p}\,'_{Db}
\quad \mbox{and} \quad
E_D = \sum_{b=1}^{n_D} E'_{Db} - E_{DP} \,.  
\end{equation}
The Fourier transform of $\psi$ is denoted by $\widetilde{\psi}$.
Note that $\vec{x}_1$ is identical with the source position
$\vec{x}_S$ for DIF. One can show for DAR with a wave function of the
neutrino source analogous to the detector wave function (\ref{psiD})
that integration over $\vec{x}_1$ causes the replacement of
$\vec{x}_1$ by $\vec{x}_S$ in the expression (\ref{x2int}) \cite{GS96}.

\subsection{The integration over $d^3 q$ and the asymptotic limit 
$L \to \infty$}
\label{theorem}

The next step in treating the expression (\ref{x2int}) is obviously
the integration over $d^3 q$. Actually, we need the result of this
integration only in the asymptotic limit of macroscopic distance
between source and detector $L = |\vec{x}_D - \vec{x}_S|$. This limit
is handled by the following theorem \cite{GS96}. \\[1mm]
\textbf{Theorem:} Suppose we have an integral of the form
\begin{equation}\label{J}
J(\vec{L}) = \int d^3 q\, \Phi(\vec{q}\,)
\frac{e^{-i\vec{q} \cdot \vec{L}}}{A - \vec{q}\,^2 + i\epsilon}\,,
\end{equation}
where $\Phi$ is three times continuously differentiable and 
$\Phi$, $\nabla_j \Phi$, $\nabla_j \nabla_k \Phi$ all decrease at
least like $1/\vec{q}\,^2$ for $L \equiv |\vec{L}| \to
\infty$. Furthermore, $A$ is a constant, i.e., independent of
$\vec{q}$ and $\vec{L}$. Then in the asymptotic limit $L \to \infty$
the integral (\ref{J}) is given by
\begin{eqnarray}
&& J(\vec{L}\,) = -\frac{2\pi^2}{L} 
\Phi \left( -\sqrt{A} \frac{\vec{L}}{L} \right) e^{i\sqrt{A}L}
+ \mathcal{O} \left( L^{-3/2} \right) \quad \mbox{for} \quad A>0 \,,
\\
&& J(\vec{L}\,) = \mathcal{O} \left( L^{-2} \right)
\quad \mbox{for} \quad A<0\,.
\end{eqnarray}

Applying this theorem to the integration 
$\int d^3 q e^{-i\vec{q} \cdot \vec{L}} \cdots$ in the expression
(\ref{x2int}), for each $j$ we identify $A$ with $(q^0)^2 - m^2_j$ and 
$\vec L$ with $\vec{x}_D - \vec{x}_S$. 
Apart from a factor, the result of the integration and 
the asymptotic limit $L \to \infty$,
which picks out the term proportional to $L^{-1}$, is described by the
following operations on the right-hand side of Eq.~(\ref{x2int})
for each $j$: 
\begin{enumerate}
\item
The denominator in the neutrino propagator is removed.
\item
In the three places in expression (\ref{x2int}) where $\vec{q}$ occurs,
i.e., in the exponential, in the numerator of the neutrino propagator 
and in $\widetilde{\psi}_D$, the replacement
\begin{equation}
\vec{q} \to - q_j \vec{\ell}
\quad \mbox{with} \quad q_j = \sqrt{E_D^2 - m^2_j}
\quad \mbox{and} \quad \vec{\ell} = (\vec{x}_D - \vec{x}_S)/L
\end{equation}
has to be made.
\end{enumerate}
This amounts to having neutrino $j$ on mass shell with 
four-momentum~\cite{GS96}
\begin{equation}\label{kin}
k_j = 
\left( \begin{array}{c} E_D \\ q_j \vec{\ell} \end{array} \right) \,.
\end{equation}
Note that with $k_j$ the numerator of the neutrino propagator is
trivially rewritten as
\begin{equation}\label{projector}
-\not\! k_j + m_j = -\sum_{\pm s} v_j(k_j,s) \bar{v}(k_j,s) \,,
\end{equation}
which is the projection operator on the negative-energy solutions of
the free Dirac equation. Of course, due to the left-handedness of the
weak charged currents the term with negative helicity is negligible in
Eq.~(\ref{projector}). 

This consideration conforms to our physical intuition: if we have a
production process for a particle with which one performs a scattering
experiment at a distance macroscopically separated in space 
from the production process, then the amplitude of the combined
production -- scattering (detection) process is the product of the
production and scattering amplitudes. This is the physical content of
the above theorem. In the case of neutrino oscillations the particles
with which the scattering is performed are the massive
neutrinos. 

In the case of neutrino mixing (\ref{mixing}) several
different massive neutrinos are produced and the amplitude is the sum
over the products of production and detection of the neutrino mass
eigenstates with mass $m_j$. In this way interference effects arise if
certain conditions, which will be explored in the following, are fulfilled. 
In summary, after having
performed the integrations over $d^4 x_2$ and $d^4 q$, 
in the asymptotic limit
$L \to \infty$ only the neutrinos on mass shell contribute to the
amplitude in Eq.~(\ref{Pav}) which can be written as~\cite{GS96,GSM99}
\begin{equation}\label{ampinfty}
\mathcal{A}^\infty_{\nu_\alpha\to\nu_\beta} =   
\sum_j \mathcal{A}^S_j \mathcal{A}^D_j
U_{\beta j}U^*_{\alpha j} e^{i q_j L} \,,
\end{equation}
where
$\mathcal{A}^S_j$ and $\mathcal{A}^D_j$ denote the amplitudes for 
production and detection of a 
neutrino with mass $m_j$. Note that $E_D$ is the energy on the neutrino line
in Figs.~\ref{DAR} and \ref{DIF}. It is independent of $m_j$ and
determined by the energies of the final states of the detection
process. This is an immediate consequence of the assumptions in the 
previous section, in particular, of Assumption I. Furthermore,
due to the above-mentioned integrations and the asymptotic limit we obtain
\begin{equation}\label{AD}
\mathcal{A}^D_j \propto \widetilde{\psi}_D(-q_j \vec{\ell} + \vec{p}\,'_D)
\,.
\end{equation}
In the case of DAR an analogous function $\widetilde{\psi}_S$
appears ($\psi_S$ is localized at $\vec{x}_S$), such that~\cite{GS96,GSM99}
\begin{equation}\label{AS}
\mathcal{A}^S_j \propto \widetilde{\psi}_S
(q_j \vec{\ell} + \vec{p}\,'_S )
\quad \mbox{with} \quad
\vec{p}\,'_S = \sum_{b=1}^{n_S} \vec{p}\,'_{Sb} \,.
\end{equation}

The amplitude (\ref{ampinfty}) looks very similar to the amplitude
appearing in the standard oscillation formula (\ref{P}), and the
question when the standard formula arises amounts to a discussion of
the conditions under which the amplitudes $\mathcal{A}^S_j$ and 
$\mathcal{A}^D_j$ are independent of $j$. This will be done in the
next section.

In Ref.~\cite{GSM99} it has been shown that the finite lifetime of the
neutrino source particle can be incorporated with the help of the
Weisskopf--Wigner approximation, thus performing perturbation theory
starting at a final initial time $t_i = 0$. Consequently, in the
$d^4 x_2$ integration, the time variable $x^0_2$ is integrated from 0 to
$\infty$ and, therefore, the $dq^0$ integrand is not simply given by
$\delta (q^0 + E_D)$ (see Eq.~(\ref{x2int})). Nevertheless, it has been
demonstrated in Ref.~\cite{GSM99} that in the limit of macroscopic $L$
one recovers this delta function and all our manipulations at the
detector vertex and the inner neutrino line discussed in this section
are valid also for explicit finite lifetime of the neutrino
source. However, there are effects stemming from the source
vertex. These will be discussed in the next section without
derivations. For details we refer the reader to Refs.~\cite{GSM99,GMS99}.

\section{Results}

The preceding discussion based on the assumptions
stated in Section \ref{assumptions} leads us to the following conclusions:
\begin{enumerate}
\renewcommand{\labelenumi}{(\roman{enumi})}
\item
Since in the asymptotic limit $L \to \infty$ the neutrinos on the inner line in
the Feynman diagrams Fig.~\ref{DAR} and \ref{DIF} are on mass shell we are
allowed to make the interpretation that we have neutrino mass eigenstates 
$| \nu_j \rangle$ characterized by the energy $E_\nu \equiv E_D$ 
$\forall j$ and momenta $q_j = \sqrt{E_D^2 - m_j^2}$.
Thus they have all the \emph{same energy} determined by the 
\emph{detection process}, but the momenta are different \cite{GSM99}. The same
conclusion has also been reached in~\cite{lipkin,grossman}, 
however, for other reasons.
\item
The summation over $E_D$ is incoherent, i.e., it occurs in the cross section
(see Eq.~(\ref{Pav})), not in the amplitude (\ref{ampinfty}). In this sense
there are no neutrino wave packets\footnote{The same conclusion but
for a different reason was presented in Ref.~\cite{stodolsky}.} 
in experiments conforming to our assumptions \cite{GSM99}. 
\item
It has been shown in Ref.~\cite{KNW96} that the relation (\ref{Lcoh}) 
holds irrespective of the
coherent or incoherent nature of the energy spread. However, with the two
points just made it follows that any coherence length originates in an
incoherent energy spread and reflects the inability to measure the energies of
the final states of the detector process more precisely than 
$\Delta E_\nu$ \cite{GSM99}. Note that in~\cite{KNW96} it has already
been pointed out that a coherent or incoherent neutrino energy
spread cannot be distinguished in neutrino oscillation experiments.
Since the neutrino energy can in principle be determined
with arbitrary precision the coherence length can theoretically be increased
solely by detector manipulations \cite{KNW96,GK98}.
\item
From Eq.~(\ref{AD}) it follows that with 
$\Delta m^2_{kj} \equiv m^2_k-m^2_j$ ($k>j$)
the condition
\begin{equation}\label{ACC}
q_j-q_k \simeq \frac{\Delta m^2_{kj}}{2E_D} 
\lesssim \sigma_D \quad \mbox{or} \quad
\sigma_{xD} \lesssim \frac{1}{4\pi} L^\mathrm{osc}_{kj}
\quad \mbox{(DAR + DIF)}
\end{equation}
is necessary for having neutrino oscillations with 
$\Delta m^2_{kj}$ (see Eq.~(\ref{PP})),
where $\sigma_D$ and $\sigma_{xD}$ are the widths of the wave function of the
detector particle in momentum and coordinate space, respectively. Otherwise,
the standard formula (\ref{P}) has to be corrected by suppressing the
oscillatory term containing $\Delta m^2_{kj}$. 
The second part of equation (\ref{ACC}) is obtained 
by rewriting the first part using the
oscillation length (\ref{Losc}) \cite{kayser,rich,GS96}. In
realistic experiments condition (\ref{ACC}) holds because $\sigma_{xD}$ 
is a microscopic quantity, whereas, dropping from now on the indices
$j$, $k$, the oscillation length
\begin{equation}\label{Lmacr}
L^\mathrm{osc} \simeq 2.48 \: \mbox{m} \: 
\left( \frac{1 \: \mathrm{eV}^2}{\Delta m^2} \right)
\left( \frac{E_\nu}{1 \: \mathrm{MeV}} \right)
\end{equation}
is a macroscopic quantity. 
\item
In the case of DAR a condition analogous to (\ref{ACC})
\begin{equation}\label{ACCS}
\sigma_{xS} \lesssim \frac{1}{4\pi} L^\mathrm{osc}
\quad \mbox{(DAR)}
\end{equation}
exists for the width of
the neutrino source wave function. Note that the conditions
(\ref{ACC}) and (\ref{ACCS}) are
independent of $L$, i.e., Eqs.~(\ref{ACC}) and (\ref{ACCS}) do 
not define a coherence length. Such conditions were first discussed in
the wave packet approach by Kayser \cite{kayser}.
\item
Now we want to consider effects of the finite lifetime 
$\tau_S = 1/\Gamma$ of the neutrino source. The decay width
$\Gamma$ introduces an energy spread $\Delta E_\nu \sim
\Gamma$. However, the corresponding coherence length (\ref{Lcoh}) is
much too long to be of interest in realistic experiments if we take
the decay widths of muons or charged pions. If the neutrino source
particle can be conceived as a free, unbound particle its wave
functions spreads during its lifetime. It has been shown~\cite{GSM99}
that, in order not to wash out neutrino oscillations, the condition
\begin{equation}\label{SFC}
\frac{\Delta m^2 \sigma_S}{m_S E_D} \lesssim \Gamma
\end{equation}
has to be fulfilled, where $\sigma_S$ is the spread of
$\widetilde{\psi}_S$. Defining a velocity spread of the source
particle at rest by $\sigma_S/m_S$, where $m_S$ is its mass, we can
rewrite Eq.~(\ref{SFC}) as
\begin{equation}
\Delta v_S\, \tau_S \lesssim \frac{1}{4\pi} L^\mathrm{osc} \,.
\end{equation}
We can interpret this relation by saying that the spreading of the
source wave function during the lifetime of the source particle
should be less than the oscillation length in order not to destroy 
the oscillation pattern.
\end{enumerate}

Some comments concerning these results are at order. The points (i) --
(iv) refer to only to the detector. In particular, (i) -- (iii) mean
that the nature of the neutrino energy spread is determined by the
properties of the detector. This is an immediate consequence of the
Assumptions I, II, III in Section \ref{assumptions}. Making a simple
numerical estimate of Eq.~(\ref{SFC}) for the LSND~\cite{LSND} 
and KARMEN~\cite{KARMEN}
experiments, where the source particles are stopped muons, we have
$\sigma_S \lesssim 0.01$ MeV~\cite{private}, $\Delta m^2 \sim 1$ eV$^2$
and $E_\nu \sim 30$ MeV. Thus we get 
$\Delta m^2 \sigma_S/m_\mu E_\nu \lesssim 3 \times 10^{-18} \;
\mbox{MeV} \ll \Gamma_\mu \approx 3 \times 10^{-16}$ MeV. Therefore, the
coherence condition (\ref{SFC}) is very well fulfilled (see
Ref.~\cite{GMS99} for a detailed derivation and discussion of this condition). 
The quantum-mechanical conditions (\ref{ACC}) and (\ref{ACCS})
have classical analogues: the uncertainty in the position of 
the detector (source) particle has to be less than the oscillation
length, otherwise oscillations are washed out by these macroscopic,
classical averaging processes \cite{pon}. Consequently, if these classical
conditions, which involve macroscopic lengths, are fulfilled, then
the conditions (\ref{ACC}) and (\ref{ACCS}) involving the
microscopic spread of wave functions hold to a much better degree.
In summary, in the framework discussed here all corrections to
Eq.~(\ref{P}) are negligible.

\section{What remains to be discussed}

It remains to point out what is still missing or what was not
discussed in the field-theoretical approach. In the case of DIF,
conditions concerning the target process (replacing conditions
concerning the source process for DAR) have to be worked out. For an
attempt in this direction see Ref.~\cite{campagne}. 
Note that we have not been able to build in
the interaction of the final state particles in the source with the
environment. This is a potential source of a finite coherence
length~\cite{nussinov,KNW96} according to the following consideration:
neutrino emission is interrupted at the time when one of the final
particles in the source process interacts with the environment. This
leads to a finite length $\ell_\nu$ of the neutrino wave train and
thus to an energy spread $\Delta E_\nu \propto 1/\ell_\nu$, resulting
in a finite coherence length~(\ref{Lcoh}). Such a coherence length
might play a role for solar neutrinos with
$\Delta m^2 \sim 10^{-5}$ eV$^2$ (MSW effect~\cite{MSW}), in which
case the mass-squared difference and $L = 150 \times 10^6$ km
could be large enough that the neutrino
wave packets corresponding to the mass eigenstates are separated when
they arrive on earth \cite{nussinov,KNW96}. However, the 
different averaging processes involved in the calculation of
the solar neutrino flux lead in any case
effectively to the consideration of
the neutrino state arriving on earth as an incoherent mixture of
neutrino mass eigenstates (see, e.g., Ref.~\cite{dighe}). A further
problem is the inclusion of the MSW effect in the field-theoretical approach.
This problem was studied in Ref.~\cite{cardall}. In the theorem
discussing the asymptotic limit $L \to \infty$ (see Section
\ref{theorem}) we have assumed that $L$ pertaining to a neutrino
oscillation experiment is so large that in the specific situation $L$
is in the asymptotic region of the integral (\ref{x2int}). It was found 
in Ref.~\cite{ioan} that under certain conditions there is an
intermediate range for $L$ which results in an oscillation formula
different from (\ref{P}). There is also the possibility
that in some cases (e.g., the KARMEN experiment) it is not realistic to 
use the conventional procedure to calculate the cross section (\ref{Pav}) by
taking the asymptotic limit of the final time to infinity.

\section{Concluding remarks}

The field-theoretical treatment of neutrino oscillations includes the
neutrino production and detection processes and allows thus a
description in terms of quantities which are really manipulated in an
experiment. This treatment has demonstrated the robustness of the
canonical oscillation formula (\ref{P}) against corrections in
realistic experimental situations. The reason is that in the
field-theoretical approach, for the
validity of Eq.~(\ref{P}), only the conditions 
(\ref{ACC}), (\ref{ACCS}) and (\ref{SFC}), which
do not involve $L$, have to be fulfilled. 
The first two conditions simply state that the
macroscopic oscillation length (\ref{Lmacr}) has to be much larger
than the microscopic spatial extension of the wave functions of the
neutrino detection and source particles. Condition (\ref{SFC}) refers
to unstable neutrino sources and seems to be well fulfilled in view
of~\cite{LSND} $\Delta m^2 \lesssim 2$ eV$^2$ for all realistic
sources. Conditions involving $L$ (oscillation lengths) stemming from
interactions of final state particles of the source process with
surrounding matter have not yet been included in the field-theoretical
approach. However, we do not expect that their inclusion drastically
changes the heuristic estimates of coherence lengths. 

A theoretical
aspect of the approach presented here is that the properties of the
oscillating neutrino state is largely determined by the detector.
With the reasonable assumptions introduced in Section
\ref{assumptions} we arrive at the conclusion that no
neutrino wave packets are needed for the description of neutrino
oscillations. In some sense, neutrino wave packets are replaced by the
wave functions of the neutrino detector and source (target) particles.
Using neutrino flavour states
\begin{equation}\label{state}
| \nu_\alpha;t,x \rangle = \sum_j U_{\alpha j}^* | \nu_j \rangle
\, e^{-i(E_\nu t - q_j x)}
\quad \mbox{with} \quad q_j = \sqrt{E_\nu^2 - m_j^2}
\end{equation}
and $\langle \nu_j | \nu_k \rangle = \delta_{jk}$
in the discussion of neutrino oscillations
and performing neutrino energy summations in the probabilities
(\ref{P}) but not in the states (\ref{state}) comes
close to the spirit of the field-theoretical approach.

\section*{Acknowledgments}
W.G. would like to thank W. Alberico for organizing the meeting in
honour of Samoil Bilenky's 70th birthday.

\vspace{1cm}

\end{document}